\documentstyle[sprocl]{article}

\begin{document}

\title{Extraction of skewed parton distributions.
\vspace{1cm}}

\author{Dieter M\"uller\vspace{1cm}} 

\address{Institut f\"ur Theoretische Physik, Universit\"at Regensburg,\\
D-93040 Regensburg, Germany\vspace{1cm}}

\maketitle\abstracts{Skewed parton distributions contain new
non-perturbative information about hadronic states. Thus, their extraction
from experimental data is an important goal. Properties and models for
skewed parton distributions as well as their extraction, based on
perturbative leading-order results, are shortly reviewed.}

\vspace{5cm}

{\noindent\it Talk given at the 8th International Workshop on Deep
Inelastic scattering and QCD, Liverpool, April 25-30, 2000.}

\newpage

\section{Definition and properties of Skewed parton distributions.}

Skewed parton distributions (SPD's) were introduced for about one
decade as a generalization of both Feynman's parton densities as well as of
exclusive distribution amplitudes \cite{DVCS}. They are defined
as expectation values of light-ray operators sandwiched between hadronic
states $| P_i S_i \rangle$ with different momentum $P_i$ and spin $S_i$
dependence. At twist-two level they read in the quark sector:
\begin{eqnarray}
\label{Def-SPD}
\left\{ { {^Q\!q^V} \atop {^Q\!q^A} } \right\} ( x, \xi,\Delta^2,\mu)
= \int \frac{d\kappa}{2\pi} e^{i \kappa x P_+}
\langle P_2 S_2 |
\bar \psi (- \kappa n)
\left\{ { \gamma_+ \atop \gamma_+ \gamma_5 } \right\}
\psi (\kappa n)
| P_1 S_1 \rangle_{\mu} ,
% \\
%\left\{ { {^G\!q^V} \atop {^G\!q^A} } \right\} ( t, \xi,\Delta^2,\mu^2)
%&=& \frac{4}{P_+} \int \frac{d\kappa}{2\pi} e^{i \kappa t P_+}
%\langle P_2 S_2 |
%G^a_{+ \mu} (-\kappa n)
%\left\{ { g_{\mu\nu} \atop i \epsilon_{\mu\nu-+} } \right\}
%G^a_{\nu+} (\kappa n)
%| P_1 S_1 \rangle_{\mu},
\end{eqnarray}
where $x$ is the longitudinally momentum fraction with respect to
$P_+=n(P_1+P_2)$ [$n$ is a light-cone vector which project onto the +
component], $\xi=-\Delta_+/P_+$ is the so-called skewedness parameter
with $\Delta=P_2-P_1$ and $\mu$ is the renormalization
scale of the operators. They describe the probability amplitude to find a
quark with momentum fraction $(x-\xi)P_+/2$ which goes for the
DGLAP--region, i.e.\ $|x|\ge |\xi|$, into a final quark with momentum
fraction $(x+\xi)P_+/2$ or forms together with the second quark a
mesonic like state into the ER-BL--region, i.e.\ $|x|\le |\xi|$.

More recently, it has been realized that they contain valuable
non-perturba\-tive information, which, for instance, may offer the
possibility to measure the angular momentum fraction of quarks and gluons in
the nucleon. A number of properties follow from first
principals by means of the definition (\ref{Def-SPD}):
\begin{itemize}
\item Support properties: $ q(x,\xi)=0$ for $|x|> 1$.
\item Polynomial property of moments:
 $ \int_{-1}^1 dx x^j q(x,\xi,\Delta^2)= \sum_{k=0}^j q_{jk}(\Delta^2)\xi^k$.
\item Lowest moment is given by form factors:
$ q_{00}=
% \int_{-1}^1 {^Q\!q^I}(x,\xi)
\langle P_2 S_2 | n^\mu {^Q\!J^I_{\mu}} | P_1 S_1 \rangle/P_+.$
\item Hermiticity as well as time reversal invariance provide:
      $q(x,\xi)= q(x,-\xi)$.
\item Evolution equation follows from renormalization group invariance:
\\
$ \mu^2\frac{d}{d \mu^2} q(x,\xi,\Delta^2,\mu)=\int_{-1}^1\frac{dy}{|\xi|}
V(x/\xi,y/\xi;\alpha_s(\mu))q(y,\xi,\Delta^2,\mu)$.
\item Inclusive connection: $q(x,\mu^2)=
     q(x,\xi=0,\Delta^2=0,\mu^2)_{|S_2 \to S_1}$,
where $q(x,\mu^2)$ is the usual parton density.
\item Exclusive connection:
$\Phi(x,\mu^2)=
     q(x,\xi=-1,\Delta^2=M^2,\mu^2)_{\langle P_2,S_2| \to \langle 0|}$,
where $|P_1,S_1\rangle$ is now a meson state and  $\Phi$
is the distribution amplitude.
\end{itemize}

\section{Models for skewed parton distributions.}

As we see from the list given above, the SPD's give us a link between
exclusive quantities like electromagnetic form factors and parton densities
measured in inclusive reactions. Based on model assumptions, there exist
different proposals for the SPD's with quite different characteristics. For
instance, the bag-model predicts a valence quark distribution which is
rather independent on the skewedness parameter \cite{JiMelSon97}. While the
chiral solution model in the large $N_c$ limit takes into account the Dirac
sea and, thus, predicts a more complex shape of SPD's containing zeros and a
strong skewedness dependence \cite{PetPobPolBoeGoeWei97}. Note that the
typical sclale for these predictions is very low, i.e.\ $Q_0\approx 0.4$ GeV
and $Q_0\approx 0.6$ GeV, respectively.

Other suggestions are inspired by the inclusive connection and are based on
different mappings of the parton densities to the SPD's, for instance,
equating them, mapping the Mellin moments to the conformal ones due to an
integral transformation, or expressing the SPD's in terms of double
distributions \cite{furpro}. Thereby, it is assumed that the $\Delta^2$
dependence is factorized, which is certainly true up to logarithmic
corrections for large $\Delta^2$ of a few GeV. Note that in the forward
limit the spin flip part vanishes, and therefore, further assumptions are
needed. For very small values of $\xi$, the non-spin flip part in the DGLAP
region is essentially determined by the forward parton distributions.

These prescriptions have to be supplemented by the scale at which they are
applied. It seems to be a good idea to use a low input scale that is
typically for non-perturbative model calculations. Afterwards one evolves
the SPD's to the scale that is used in hard scattering experiments, i.e.\
higher than at least one GeV. The advantage is twice: i) one study the
stability under evolution and ii) perturbative QCD information are taken
into account. Note that evolution to an asymptotic large scale predicts,
independently on the initial conditions, that the SPD's are concentrated in
the ER-BL region and vanish for $x=\pm \xi$. However, this fact is not of
practical relevance for the scales accessible in experiments.

\section{Experiments to access skewed parton distributions.}

There is a growing interest to access these non-perturbative functions in
lepton-hadron scattering experiments, where the virtuality of the
intermediate photon has to be larger than of at least one GeV. On the
theoretical side there are formal factorization proofs for the following
hard processes to leading twist-two accuracy in a perturbative calculable
hard-scattering part and universal SPD's: deeply virtual Compton scattering
(DVCS) $e^\pm p \to e^\pm p \gamma$, exclusive meson $M$ production $e^\pm p
\to e^\pm B M$\footnote{ Since $B$ denotes an arbitrary baryon, the SPD's
defined in eq.\ (\ref{Def-SPD}) are generalized to ``off-diagonal'' ones.
Note also that factorization is only proofed for longitudinal polarized
photon.}, and exclusive lepton pair production $e^\pm p \to e^\pm p l^+
l^-$. In the first two cases the amplitude reads in leading order:
\begin{eqnarray}
\label{Pre-Pro}
{\cal A}(\xi,\Delta^2,Q^2) \propto
\sum_{j=u,d,s}\int_{-1}^1 dx \left(\frac{C_j}{x-\xi+i\epsilon}
+ \frac{\overline{C}_j}{x+\xi-i\epsilon} \right)
q_j(x,\xi,\Delta^2,Q^2),
\end{eqnarray}
where the coefficients $C_j,\overline{C}_j$ depend on the considered
process. If ${\cal A}$ could be very precisely measured as function of $\xi$
and $Q^2$, the deconvolution of this formula exists in principal.
Practically, one has to compare the model predictions with the experimental
data or defines characteristic functions that together with eq.\
(\ref{Pre-Pro}) can distinguish between different models:
%For instance, the ratio
%$R={\rm Re} {\cal
%A}(\xi,\Delta^2,Q^2)/ {\rm Im} {\cal A}(\xi,\Delta^2,Q^2)$ or the integral
\begin{eqnarray}
R=\frac{{\rm Re} {\cal
A}(\xi,\Delta^2,Q^2)}{ {\rm Im} {\cal A}(\xi,\Delta^2,Q^2)},\quad
S = 1 - \frac{{\rm PV}\int_{-1}^1 dx \frac{1}{x-\xi}
%\mp \frac{1}{x+\xi}\right)
{\rm Im} {\cal A}(x,\Delta^2,Q^2)
}{(-\pi) {\rm Re} {\cal A}(\xi,\Delta^2,Q^2)}.
\end{eqnarray}
The latter one can be considered as a measure for the skewedness dependence.

Unfortunately, this perturbative leading order analysis can be spoiled by
the size of perturbative as well as higher twist corrections. The first ones
have been considered for DVCS in the valence quark region at next-to-leading
order. It turns out that the corrections due to evolution are small, i.e.\
about 10\% or less, while the corrections to the hard scattering amplitude
depend on the chosen model and can be of the order of 50\% or even more
\cite{NLOcor}.

%\bibliography{c:/user/TeXinput/referenc}
%\bibliographystyle{unsrtRev}

\end{document}